\documentclass{nature}

\usepackage{hyperref}
\usepackage[pdftex]{graphicx,color}
\usepackage{amsmath, amssymb}
\usepackage{graphicx}
\usepackage[english]{babel}
\usepackage{subcaption}
\usepackage{enumerate}
\usepackage{tikz}
\usetikzlibrary{positioning}

\graphicspath{{./figs/}}

\newcommand{\aap}{Astron.~Astrophys.}

\let\oldbibliography\thebibliography
\renewcommand{\thebibliography}[1]{\oldbibliography{#1}
\setlength{\itemsep}{-5pt}}

\title{Reply to: Overconfidence in Bayesian analyses of galaxy rotation curves}

\author{Davi C. Rodrigues$^{1,2}$, Valerio Marra$^{1,2}$, Antonino Del Popolo$^{3,4,5}$ \&  Zahra Davari$^{6}$}

\begin{document}

\maketitle

\begin{affiliations}
 \item {\footnotesize Center for Astrophysics and Cosmology, CCE,  Federal University of Esp\'irito Santo, 29075-910, Vit\'oria, ES, Brazil.}
 \item {\footnotesize  Department of Physics, CCE,  Federal University of Esp\'irito Santo, 29075-910, Vit\'oria, ES, Brazil.}
 \item {\footnotesize  Dipartamento di Fisica e Astronomia, Universit\`a di Catania, Viale Andrea Doria 6, 95125 Catania, Italy.}
 \item {\footnotesize  Institute of Astronomy, Russian Academy of Sciences, 19017, Pyatnitskaya str., 48 , Moscow.}
 \item {\footnotesize  INFN sezione di Catania, Via S. Sofia 64, 95123 Catania, Italy.}
 \item {\footnotesize  Department of Physics, Sharif University of Technology, P.O.Box 11365-9161, Tehran, Iran.}
\end{affiliations}

\bigskip

Cameron {\it et al.}~2019\cite{2020NatAs.tmp...15C} (hereafter C19) recommends a more cautious and rigorous approach to statistical analysis in astronomy. We welcome this particular side of their communication as it helps stimulating the effort towards the adoption of better statistical methods in galaxy rotation curves, an effort to which we  contributed with Rodrigues {\it et al.}~2018\cite{Rodrigues:2018duc} (hereafter R18). Indeed, R18 was the first work that, in order to conclude on the universality of the acceleration scale $a_0$, studied the posterior distributions on $a_0$ of a large set of galaxies. As C19 agrees, the credible intervals were found within the Bayesian framework, that is, the  marginalized posteriors on $a_0$ were found using Bayes' theorem to update the priors in light of the observational data; this process was done without introducing any approximation. Considering R18, C19 also remarks that: i) better methods to select the nuisance parameters and the corresponding priors could be used; ii) a quality cut based on $\chi^2$ values should not be used, and iii) the compatibility of the posteriors should be assessed in a more robust way. In the following, after first clarifying the context of our work, we address these criticisms.

\noindent
\subsection{Context.} There are several theories for galaxy dynamics in which a universal parameter such as $a_0$ is introduced and whose value is {\it a priori} unknown, MOND being one example. A common practice\cite{Martins:2007uf, Gentile:2010xt} is to perform fits of several galaxies and compute the average of the best fits in order to find the ``best'' universal value. This procedure does not fully exploit the data as only the peak of the likelihood is considered, nor it tests whether data is compatible with the hypothesis that such a universal parameter exists. McGaugh {\it et al.}~2016\cite{McGaugh:2016leg} used another approach: the supposed universal parameter $a_0$ is found by fitting a phenomenological function to all the data from all the galaxies of the sample at the same time. Since the rms scatter of the residuals was found to be similar to the total expected rms due to observational errors, this was interpreted as suggesting that there is no, or very little, $a_0$ intrinsic variance among different galaxies. This is an interesting approach, with different possible applications, but it is not sensitive to test the existence of a fundamental acceleration scale. This context motivated our work. Some previous tests on the universality of $a_0$  with other samples, less galaxies  and without computation of the (Bayesian) credible intervals can be found in the literature\cite{Randriamampandry:2014eoa}. We investigated this issue  within the Bayesian framework and  performed a type of test that is frequently performed in a cosmological context (such as the compatibility between BAO and CMB) but not in the context of galaxy rotation curves.

\subsection{Nuisance parameters and priors.} In the context of galaxy rotation curves, the discussion regarding the set of nuisance parameters and their priors did not yet reach the level of sophistication common in cosmological analyzes that have been using Bayesian inference since the early 90's. Specifying the priors is well known to be a subtle task for several reasons. Regarding R18, we remark that three very different sets of priors for the nuisance parameters were explored by R18 (in their Fig.~1), Rodrigues {\it et al.}~2018\cite{Rodrigues:2018lvw} (Fig.~2.a) and Chang \& Zhou 2019\cite{Chang:2018lab} (Fig.~4.a). One concludes that priors that include the SPARC reference values preserve the distribution of the $a_0$ modes, which span almost two orders of magnitude if $a_0$ is given an uninformative prior, which is justified as the value of $a_0$ is {\it a priori} unknown. Therefore, there will be compatibility among the posteriors on $a_0$ if the widths of their credible intervals are sufficiently large, and this does depend on the choice of the priors. Chang \& Zhou~2019\cite{Chang:2018lab} used  the same priors of Li {\it et al.}~2018\cite{Li:2018tdo} for the nuisance parameters, and found that the $a_0$ credible intervals were on average enlarged with respect to R18. Nonetheless, they confirmed that, using the compatibility criteria of R18, a fundamental acceleration is rejected at more than 10$\sigma$ (in qualitative agreement with quality-cut based arguments\cite{Rodrigues:2018lvw}). C19 lists several possible ways to improve the choice of the nuisance parameters and priors within Bayesian inference, and we agree that some of the procedures should be pursued considering applications to galaxy rotation curves in general.

\subsection{Quality cuts based on $\chi^2$.} C19 stresses that we motivated the use of this quality cut using expectations outside Bayesian inference. This quality cut was motivated considering that exceptionally high values for the minimum $\chi^2$  (outside the expected $5 \sigma$ level for $\chi^2$-statistics) would indicate a dubious fit (i.e., if one assumes the model is right, then there should be an issue with the observational data), and thus such case could lead to a less trustworthy determination of the $a_0$ posterior for that galaxy. Although the problematic nature of quantitative analyzes based on the $\chi^2$ is well known, it is still useful as a {\it guidance} regarding model performance at a qualitative level. Note that we did not use the $\chi^2$ values of the galaxies in the subsequent steps of the analysis, but rather focused on the marginalized posterior distributions on $a_0$. We also note that we provided the credible intervals for all the galaxies as supplementary information\cite{Rodrigues:2018duc}, including those that did not pass the quality cuts. Although this quality cut was not justified within Bayesian inference, our conclusions do not depend on it as it only reduces the significance of our findings at face value and is an additional safety measure against possible unknown systematics.  We agree that our analysis would be even more robust if the outliers were handled by using, for instance, non-Gaussian distributions\cite{Feeney:2017sgx}. Finally, selection criteria such as galaxy morphological statistics for identifying mergers may sound interesting, but this is not a simple or unambiguous task. Also, we point out that the galaxies from the samples that we used were already pre-selected for being significantly uniform and without a clear relevant interaction with other galaxies.

\subsection{Compatibility of the posteriors on $a_0$.}
After providing the $a_0$ credible intervals for all the galaxies, R18 analyzed their compatibility. As it is qualitatively clear from Figure 1 and Table 1 of R18, the hypothesis of a universal $a_0$ is not favored by the data---for example, there is no $a_0$ value that is inside the 5$\sigma$ regions of all the galaxies. In order to estimate the overall tension among the 100 galaxies that passed our main quality cuts, we used a Gaussian approximation for each of the galaxy posteriors and assumed a $\chi^2$ distribution, as detailed in R18. C19 criticizes this final step. We agree that this method can be improved and made more robust, but we also recall that this is a common and useful practice even in areas that use Bayesian inference more intensively. For example, it is customary to adopt this very method to quantify the tension between local and global determinations of the Hubble constant $H_0$, which reached the $4.4\sigma$ level\cite{Riess:2019cxk}. However, related to one of the suggestions by C19, the latter figure is sensitive to the tail of the adopted distribution, and  other works have employed a more robust framework, such as the Bayesian hierarchical model, in order to conclude on the significance of the claim\cite{Feeney:2017sgx}. From a pragmatical point of view, and considering the high-signal level found, the main criticism is that we have not properly modeled  the tail of the distribution and so our face-value quantitative result is not accurate. Indeed, it was not the purpose of R18 to precisely specify the level of tension between the credible intervals (as in the cosmological case above). In the main text of R18 we reported the lower significance of $10\sigma$ (instead of about $50 \sigma$ from the main analysis), which is also in agreement,  by a large margin,  with all the results with stronger quality cuts that we tested. Although this methodology could be improved, it is an advance towards more robust Bayesian inference in the context of galaxy rotation curves.

\subsection{Similar flaws.} C19 briefly states that another work\cite{Li:2018tdo} ``exhibits a number of similar flaws'', but the issues are actually quite different. Although it has other merits, its analysis on the possible variation of $a_0$ only depends on the maximum values of the likelihoods (best-fits), since it compares different models using the CDF of the reduced chi-squared. This latter approach was explicitly shown not to be robust in its context\cite{Rodrigues:2018lvw}.

\subsection{Final remarks.} The adoption of Bayesian inference is relatively new in the study of galaxy rotation curves and it is clear that the present state-of-the-art must be improved if one wants to make progress, especially in the context of the more complex models that depend on dark matter. From this point of view, the publication of C19 is timely as it will further sensibilize the community. On the other hand, we consider a couple of its remarks on R18  not in accordance with its context. At last, improvements on the analysis of raw data related to galaxy rotation curves, as briefly remarked in R18, should also be considered.

\begin{addendum}

\item DCR and VM thank CNPq and FAPES (Brazil) for partial financial support.  ZD thanks  Iran Science Elites Federation for financial support.
 
\item[Competing Interests]
The authors declare no competing financial interests.

\item[Data availability] All the relevant data for this work can be found in the cited references.

\end{addendum}

\end{document}